\documentclass[aps,amsfonts,prl,twocolumn]{revtex4}
\usepackage{amsfonts}
\usepackage{amsmath}
\usepackage{amssymb}
\usepackage{epsfig}
\usepackage{graphicx}

\newcommand{\be}{\begin{equation}}
\newcommand{\ee}{\end{equation}}
\newcommand{\ba}{\begin{eqnarray}}
\newcommand{\ea}{\end{eqnarray}}

\begin{document}
\title{$\eta/s$ and Phase Transitions}
\author{Antonio Dobado, Felipe J. Llanes-Estrada
 and Juan M. Torres-Rincon}
\affiliation{Departamento de F\'{\i}sica Te\'orica I, Universidad
Complutense de Madrid, 28040 Madrid, Spain}
\begin{abstract} 
We present a calculation of $\eta/s$  for the meson gas (zero  baryon
number), with the viscosity computed within unitarized NLO chiral perturbation theory, and 
confirm the observation that $\eta/s$ decreases towards the possible 
phase transition to a quark-gluon plasma/liquid. The value is  somewhat higher than previously estimated in LO $\chi$PT.
We also examine the case of atomic Argon gas to check the discontinuity
of $\eta/s$ across a first order phase transition.
Our results suggest employing this dimensionless number, sometimes called KSS number (in analogy with other ratios in fluid mechanics such as Reynolds number or Prandtl number)
to pin down the phase transition and critical end point to a cross-over in strongly interacting nuclear 
matter between the hadron gas and  quark and gluon plasma/liquid.
\end{abstract}
\maketitle

%%%%%%%%%%%%%%%%%%%%%%%%%%%%%%%%%%%%%%%%%%%%%%%%%%%%%%%

\maketitle

%%%%%%%%%%%%%%%%%%%%%%%%%%%%%%%%%%%%%%%%%%%%%%%%%%%%%%%%%%%%%%%%%
\section{Introduction}
%%%%%%%%%%%%%%%%%%%%%%%%%%%%%%%%%%%%%%%%%%%%%%%%%%%%%%%%%%%%%%%%%

It has been recently pointed out \cite{Csernai:2006zz} that the ratio 
of the shear viscosity to entropy density, $\eta/s$, has an extremum
 at a phase transition, based on empirical information for 
several common fluids, and follow-up calculations by us 
\cite{Dobado:2006hw} and other groups \cite{Chen:2007xe}, \cite{ChenNakano} have suggested that $\eta/s$ in a hadron gas does indeed fall slightly with 
the temperature towards the predicted transition to a quark and gluon 
plasma or liquid phase. 
Renewed interest in this quantity arose after the 
KSS conjecture \cite{Kovtun:2003wp} about a possible lower bound 
$1/(4\pi)$ (the existence of a bound had already been put forward, on
the basis of simple physical arguments, in \cite{Danielewicz:1984ww})
 and it is the subject of much current research in Heavy Ion 
Collisions. The precise reach of the bound has been under recent discussion, 
\cite{Cohen:2007zc,Son:2007xw}
and there is much interest in finding theoretical
or laboratory fluids that reach the minimum possible value of  $\eta/s$ \cite{Cohen:2007qr,Schafer:2007pr}. 

There is good hope that $\eta/s$ and even $\eta$ by itself can be derived from
particle and momentum distributions in heavy ion collisions \cite{Gavin,Lacey:2006bc,Lacey:2007na}.

  It has been shown through several examples  that, empirically,  $\eta/s$ seems to have a discontinuity at a first order phase transition, but is
continuous and has an extremum at a second order phase transition or at a crossover.

Based on lattice data \cite{Karsch:2000kv}
 it is believed that the phase 
transition between a gas of hadrons
and a quark-gluon phase at zero baryon chemical potential is actually a cross over. 
The result of \cite{Csernai:2006zz} however presents a clear discontinuity.
This is of course not a serious claim of that paper, but simply an
 artifact of the very crude approximations there employed. We here revisit the issue, 
improving as far as feasible on the hadron-side  estimate, and  further motivating the proposed behavior of $\eta/s$.

%%%%%%%%%%%%%%%%%%%%%%%%%%%%%%%%%%%%%%%%%%%%%%%%%%%%%%%%%%%%%%
\section{Inverse Amplitude Method in $\chi$PT and hadron phase
transition}
%%%%%%%%%%%%%%%%%%%%%%%%%%%%%%%%%%%%%%%%%%%%%%%%%%%%%%%%%%%%%%
We here improve the very rough calculation of
\cite{Csernai:2006zz} for $\eta/s$ on the hadron phase. We have calculated in \cite{Dobado:2003wr}
 the shear viscosity of a meson gas (that is, the
hadron gas as a function of the temperature and approximate meson chemical
potentials, at zero baryon chemical potential). That work employed the Inverse
Amplitude Method (IAM) \cite{Dobado:1989qm} that gives a good fit to the 
elastic phase
shifts for meson-meson scattering at low momentum, respects unitarity,
and is consistent with chiral perturbation theory at NLO \cite{ChPT}.
The only explicit degrees of freedom are light pseudoscalar mesons 
($\pi$, $K$, $\eta$), but elastic meson-meson resonances below 
$1\ GeV$ appear through the phase shifts \cite{GomezNicola:2001as}.

\begin{figure}
\psfig{figure=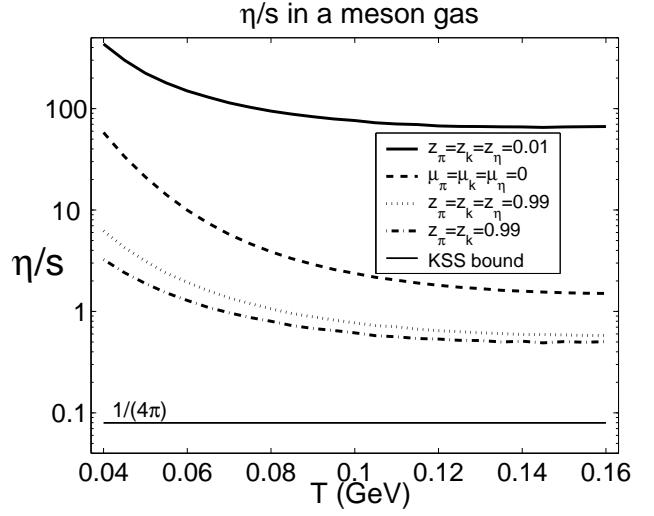,height=2.7in}
\caption{The viscosity over entropy density of a meson gas in 
chiral perturbation theory unitarized by means of the IAM. 
($z$ represents the relativistic fugacity $e^{\beta (\mu-m)}$).
\vspace{-0.3cm}
\label{fig:etameson}}
\end{figure}

\begin{figure}
\psfig{figure=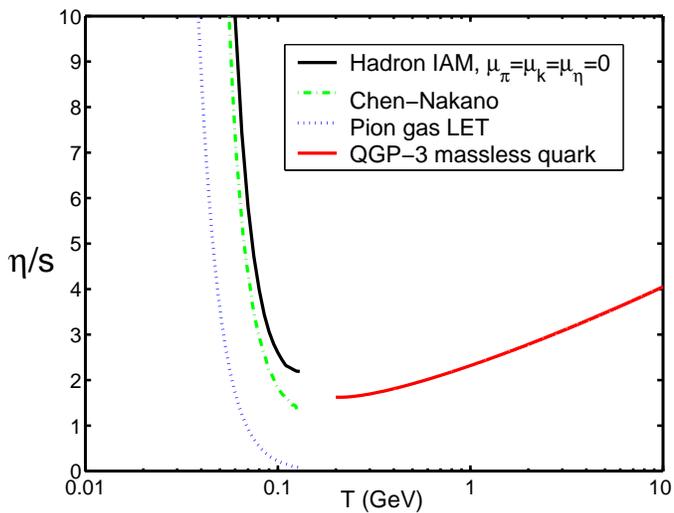,height=2.7in}
\caption{We improve the hadron-side (low $T$) estimate  of 
\cite{Csernai:2006zz} that showed the jump in the $\eta/s$ ratio in the 
transition from the hadron gas to the quark-gluon plasma, substituting 
the Low-Energy-Theorem of those authors (first order chiral perturbation 
theory) by the Inverse Amplitude Method, that agrees with Chiral 
Perturbation Theory at NLO, and satisfies elastic unitarity. We confirm the result of those authors, 
although the actual numerical value of $\eta/s$ is quite different (as 
should be expected from their calculation reaching temperatures $T\simeq
150 \ MeV$ but with only the first order interaction). One should note that,
the calculation being performed at zero baryon chemical potential, based on lattice data that suggest
a cross-over between the hadron gas and the quark-gluon plasma, and from
simple phenomenology this would suggest that $\eta/s$ should be 
continuous.
 \vspace{-0.3cm}
\label{fig:salto}}
\end{figure}

It is an elementary exercise to divide the calculated viscosity from
that work by the entropy density of the free Bose gas, for $N$ species
\be \label{sdensitymassive}
s=\frac{S}{V}=\frac{N}{6\pi^2T^2} \int_0^{\infty} dp p^4 \frac{E-\mu}{E} \frac{e^{\beta(E-\mu)}}{\left[e^{\beta(E-\mu)}-1\right]^2},
\ee
and plot the result in Fig. \ref{fig:etameson}.
Incidently, it can be seen in the figure that the holographic bound
$\frac{\eta}{s}>1/4\pi$ is not violated, which had been claimed in the literature
(we reported this in \cite{Dobado:2006hw}) but independently confirmed by \cite{FernandezFraile}. The reason is that in chiral
perturbation theory alone the cross section grows unchecked, eventually
violating the unitarity bound, which induces a very small viscosity.
Of more interest for our discussion in this work is to examine the possible
behavior across the phase transition. We take the simple estimate for $\eta/s$
in the quark-gluon plasma from  \cite{Csernai:2006zz}, but we use our
much improved calculation for the low-temperature hadron side (those authors
employ LO chiral perturbation theory without unitarization).
The result is plotted in Fig. \ref{fig:salto}. In addition we plot also the phase-shift based phenomenological calculation of \cite{ChenNakano}, that is consistent with ours but somewhat smaller.

The calculation that \cite{Csernai:2006zz} reported shows a discontinuous
jump between the QGP and the hadron gas, whereas simple-minded non-relativistic phenomenology would
make us expect a continuous function with a minimum. Our improved hadron calculation
still shows a discontinuity, although now the jump at the discontinuity
has opposite sign (our viscosity is larger since the meson-meson cross
section is smaller due to unitarity, instead of being an LO-$\chi$PT 
polynomial). Since our estimate for  $\eta/s$ is now approximate
only because of our use of the first order Chapman-Enskog expansion and
the quantum Boltzmann equation, both of which are reasonable approximations,
we feel further improvement on the hadron side will not restore
continuity, and future work needs to concentrate on evaluating the viscosity from
the QGP side.

\begin{figure}[t]
\psfig{figure=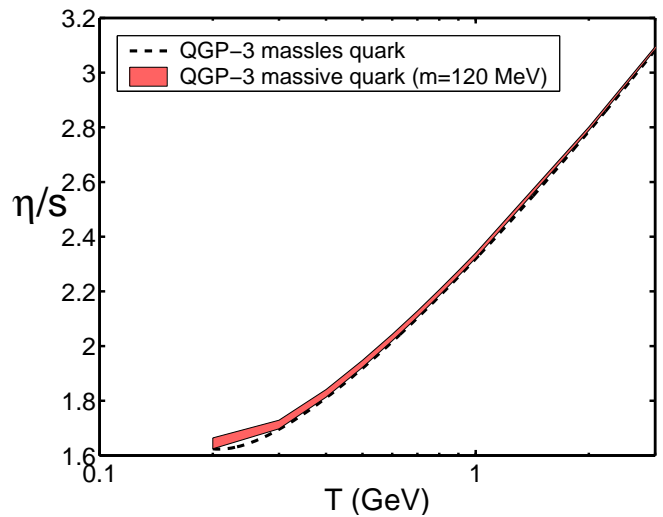,height=2.7in}
\caption{We plot the dependence of $\eta/s$ on the quark mass from quark-gluon plasma side by adapting the results from \cite{Aarts}. Note that given the non-trivial calculation there, we have slightly simplified by taking a constant $g$ in the mass correction. The band in the figure corresponds to the interval $g \in [1,2]$. We have also taken all quarks of equal mass $m_s=120 MeV$ as the maximum possible variation. As can be seen, the dependence is small and positive, bringing about even better agreement with the hadron side Inverse Amplitude Method evaluation.
 \vspace{-0.3cm}
\label{fig:saltomassive}}
\end{figure}

To calculate the viscosity in a field theory, a possible and popular approach is to employ Kubo's formula in terms of field correlators. Another method, based on the Wigner function, is to write-down the hierarchy of BBGKY equations.

In either case one can perform a low-density expansion, leading to
the use of the Boltzmann equation. Employing this on the hadron-side.
 as opposed to the full hierarchy of BBGKY equations of kinetic theory, presumes the ``molecular chaos'' hypothesis of Boltzmann, which is tantamount to neglecting correlations between sucessive collisions. This requires the collisions to be well separated over the path of the particle, and induces a systematic error in the calculation of
order $2=(m_{\pi}\lambda)$, where $1/m_{\pi}$ is the typical reach of the
strong interaction, and $\lambda$ the mean free path (controlled
by the density). To keep this number below one requires
small densities $n(T) < \frac{m_{\pi}}{2\sigma}$. If we take as a cross-section
estimate 100 $mbarn$ we see that the criterion is satisfied
up to temperatures of order 140 $MeV$ (where we stop our plot in Fig. 2).

We have also estimated the change in $\eta/s$ caused by a
small quark mass, by adapting the results of \cite{Aarts}. Those
authors provide, within a 2PI formalism, the shear viscosity
of the quark and gluon plasma of one fermion species
as a function of the fermion mass divided by the temperature,
for fixed coupling constant. Although we are
employing, as Csernai {\it et al.} do, a coupling that runs
with the scale (the temperature), the mass correction is
small, so we can take $g=2$ as fixed for a quick eyeball
estimate. We normalize the viscosity of \cite{Aarts} to the value
plotted in Figure \ref{fig:salto} at zero fermion mass, and then allow
the fermion mass to vary. The results are now plotted in
Figure \ref{fig:saltomassive}. We plot the extreme case of all three light quarks
equally massive and with mass equaling $m_s = 120 MeV$.
As can be seen, the difference to the massless case is irrelevant
at current precision and does not change the fact
that we cannot conclude as of yet whether the transition
between a hadron gas and a quark-gluon plasma/liquid
has a discontinuity in $\eta/s$ or not.
The reason that the fermion mass is not so relevant
in the calculation is twofold. First, even at $m_s$, we have
$T/m_s > 1$ for any value of $T > T_c$. Since kinetic momentum
transport in a gas is dominated by the fraction of
molecules (here, partons) with the largest energy allowed
by the Boltzmann tail of the distribution $E > T$, and the
momentum varies as
\[ p=E\left( 1-\frac{m^2}{2E^2} + \cdots\right),\]
we see that the small parton mass makes just a correction to the momentum of each (efficient) carrier. The second reason is that the cross-section, in a regime where perturbation theory is of any use, is weakly dependent on the fermion mass, with a slight dependence brought about by the logarithmic running of the quark-gluon vertex.

Still, given the large uncertainties in our knowledge of the quark-gluon
medium created in heavy-ion collisions, that make difficult to match with the hadron side, we also study
a simple non-relativistic system where the jump in $\eta/s$ at the phase transition is very clear.

%%%%%%%%%%%%%%%%%%%%%%%%%%%%%%%%%%%%%%%%%%%%%%%%%%%%%%%%%%%%%%%
\section{Liquid-gas phase transition in atomic Argon}
%%%%%%%%%%%%%%%%%%%%%%%%%%%%%%%%%%%%%%%%%%%%%%%%%%%%%%%%%%%%%%%%
In prior works it has been pointed out that experimental data 
suggest that first order phase transitions present a discontinuity in
$\eta/s$ and second order phase transitions (and maybe crossovers)
present a minimum. We will examine one case a little closer,
for a liquid-gas phase transition in the atomic Argon gas, where
we will calculate the $\eta/s$ ratio theoretically and compare to data. 
The empirical data that has been brought forward was based on atomic
Helium and molecular Nitrogen and Water. Quantum effects are very strong
in the first at low temperatures where the phase transition occurs, 
and the later have relatively strong interactions.

Instead we choose Argon due to its sphericity and closed-shell atomic structure,
that make it a case very close to a hard-sphere system. Thus, Argon
is the perfect theoretical laboratory, and sufficient data has been
tabulated due to its use as a cryogenic fluid.

The gas phase is therefore well described in terms of hard-sphere
interactions. In elementary kinetic theory one neglects
 any correlation between sucessive scatterings. The viscosity 
follows then the formula
\be \label{viscogas} 
\eta_{\rm gas} = \frac{5}{16 \, d^2} \sqrt{\frac{mT}{\pi}}\ \ ,
\ee
where $d=3.42\times10^5 \, fm$ is the viscosity diameter of the Argon atom and $m=37.3 \, GeV$ its mass
\footnote{Note that this formula follows, up to the numerical factor, from considering a classical non-relativistic gas
$\eta = \frac{1}{3} n ( m \bar{v} )\lambda$ in terms of the mean free path $\lambda$, the particle density $n$, and average momentum. The numerical factor requires a little more work with a transport equation and can be found, for example, in L. D. Landau and E. M. 
Lifshitz, ``Physical Kinetics'', Pergamon Press, Elmsford, N.Y. 1981.}. 

Experimental data is quoted as function of the temperature for fixed 
pressure. The particle density is then fixed by the equation of state;
therefore a chemical potential needs to be introduced.
In order to calculate the entropy density we again use 
Eq. (\ref{sdensitymassive}) with $N=1$.

As said, we keep the pressure $P$ constant,
and the chemical potential $\mu$ varies then within the
temperature range. In order to obtain $\mu$ we simply invert (numerically)
the function $P(T=1/\beta,m,\mu)$ at fixed temperature. 
The expression for the pressure consistent with the entropy above is
\be \label{pressuregas}
P(T,m,\mu)=-\frac{T}{2 \pi^2} \int_0^{\infty} dp \, 
p^2 \log[1-e^{\beta(\mu-E)}] 
\ee
(we have neglected in both cases the effect of the Bose-Einstein condensate
since the gas liquefies before this is relevant).
The problem has then been reduced to computing the viscosity at the given 
temperature and chemical potential, which we do employing our computer 
program for the meson gas in the Chapman-Enskog approximation, with minimum
modifications.

We change variables to absorb the scale and make the integrand of order 1 to:

\be \bar{\mu} \equiv \frac{\mu-m}{T}, \quad x \equiv \frac{p^2}{mT}.
\ee

Thus, the final expressions for the entropy density and pressure from Eqs. (\ref{sdensitymassive}) and (\ref{pressuregas}) (once integrated by parts) are

\be P=\frac{1}{12\pi^2} m^{3/2} T^{5/2} \int_0^{\infty} dx x^{3/2} \frac{1}{e^{x/2-\bar{\mu}}-1}, \ee
\ba \nonumber s_{gas}=\frac{1}{12 \pi^2} (mT)^{3/2} \int_0^{\infty} dx \, x^{3/2} \left( \frac{x}{2} - \bar{\mu} \right) \\
\times \frac{e^{\,x/2-\bar{\mu}}}{\left( e^{\, x/2-\bar{\mu}}-1 \right)^2}.  \ea

To treat the liquid Ar phase there is not a very rigorous theory. 
This is because in liquids the momentum transfer mechanism is quite complex
and does involve the interaction between molecules. Here, our choice of 
a noble gas is of help since long-range interactions are absent.
It is common to resort to semiempirical formulas with unclear theory
support, or work with formal expressions of difficult applicability. 
We compromise by combining the Van der Waals equation of state (that ultimately encodes the Lennard-Jones theory for the interatomic
potential), and use the Eyring liquid theory \cite{Eyring:1961}.

The Eyring theory is a vacancy theory of liquids. 
Each molecule composing the liquid has gas-like degrees of freedom 
when it jumps into a vacant hole, and solid-like degrees of freedom 
when fully surrounded by other molecules.

This model approach yields a partition function
 $Z$ for a one-species liquid (in natural units)

\ba \nonumber Z = \left\{ \frac{e^{E_s/N_A T}}{(1-e^{-\theta/T})^3} 
\left( 1 + n \frac{V-V_s}{V_s} e^{-\frac{aE_s V_s}{(V-V_s)N_AT}} \right) \right\}^{\frac{N_AV_s}{V}} \\
\times \left\{ \frac{e(2 \pi mT)^{3/2} V}{(2\pi)^3 N_A} 
\right\}^{\frac{N_A(V-V_s)}{V}}, \quad \ea
from which one can derive complete
statistical information about the system
\footnote{The meaning of the various variables can be found in \cite{Eyring:1961} and is as follows. $e=2.71828\dots$ is Neper's number (the presence of a single $e$ factor in the gas partition function comes from the Stirling's approximation). $E_s$ is the sublimation energy of Argon (that we express in $eV$/particle). $\theta$ is the Einstein characteristic temperature of the solid defined in any textbook. Here $a=a'$ (not to be confused with Van der Waals constant) is a model parameter, a pure-number, controlling the molecular ``jump'' between sites, or activation energy. $nV/V_s$ is the number of nearest vacancies to which an atom can jump.}.
One can recognize in the second line the partition function of a non-relativistic gas for the fraction of atoms with gas-like behavior. The first line corresponds to the solid-like behavior. The first factor is the partition function of a three-dimensional harmonic oscillator. The second term is a correction due to the translation degree of freedom, by which an atom can displace to a neighboring vacancy.

\begin{figure}
\psfig{figure=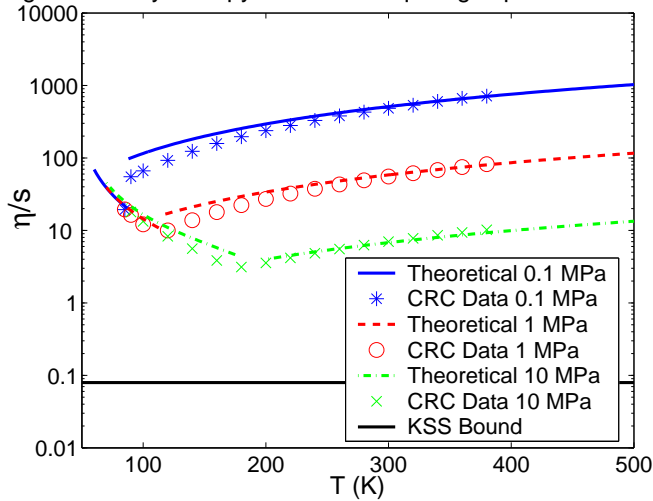,height=2.8in,angle=0}
\caption{$\eta/s$ (a pure number in natural units) for atomic Argon in 
the liquid and gas phases near the phase transition. 
Solid lines correspond to theoretical calculation described in the text, 
dashed lines are the experimental values given from \cite{CRC}. Note that
$\eta/s$ is quite independent of the pressure in the liquid phase, and that
the theoretical curves calculated from the liquid side and gas side do
get closer together with increasing pressure, suggesting as the data that 
indeed, $\eta/s$ will be continuous in the cross-over regime.
\vspace{-0.3cm}
\label{arvoe}}
\end{figure}

The shear viscosity, (like $Z$ itself), turns out to be a weighted average 
between the viscosity of solid-like (first line) and gas-like (second line) degrees of freedom 
of the liquid's particles:

\ba \nonumber \label{viscosityliquid}
\eta_{liq}=\frac{N_A 2\pi}{V} \frac{1}{(1-e^{-\theta/T})} \frac{6}{n\kappa} 
\frac{V}{V-V_s} e^{\frac{a' E_s V_s}{(V-V_s) N_AT}}  \\
+ \frac{V-V_s}{V} \frac{5}{16 \, d^2} \sqrt{\frac{mT}{\pi}}\ \ ,
\ea
and to complete the model, $\theta, n, a, a', \kappa, E_s$ and $V_s$ are given in Table \ref{table1} for gaseous Argon. $N_A$ is Avogadro's number.
The $V_s/V$ solid-like volume fraction controls the weighted average.
Note that if this ratio approaches 1, the viscosity diverges as appropriate for a rigid solid. \footnote{In this formula $\kappa$ is an ad-hoc model ``transmission coefficient''  of order 1 related to the loss of momentum to a crystal wave upon displacing an atom. Here we take it to be independent of the pressure but this could be lifted to further improve the fit in Figure \ref{arvoe}.}

\begin{table}
\caption{\label{table1} Liquid Argon parameters which appears in Eqs. (\ref{viscosityliquid}) and (\ref{entropyliquid}). All these constants are given in \cite{Eyring:1961}. However, $\kappa$ has been modified because we use Eq. (\ref{viscogas}) instead of the formula that appears in \cite{Eyring:1961} for the hard-sphere gas case.}
\begin{ruledtabular}
\begin{tabular}{cc}
Parameter & Value \\
\hline
$\theta$ & $5.17\ meV$\\
$n$ & $10.80$ \\
$a=a'$ & $0.00534$ \\
$\kappa$ & 0.667 \\
$E_s$ & $0.082\ eV/particle$\\
$V_s$ & $4.16\times10^{16}\ fm^3/particle$ \\

\end{tabular}
\end{ruledtabular}
\end{table} 

The entropy is calculated as usual taking a derivative of
the Helmholtz free energy ($A\equiv -T \log Z$),
\be  \label{entropyliquid}
 S=\frac{\partial (T\log Z)}{\partial T}\ \ . 
\ee
For our purposes we also need the liquid density which is easily 
estimated by means of the Van der Waals equation of state, that is of
some applicability in the liquid phase. 
This equation takes into account the volume excluded by the particles 
and the attraction between them. In the simplest form the Van der Waals equation is:
\be \label{vanderwaals} \left( n_{gas} + n_{liq}^2 \frac{a}{T} \right) (1-n_{liq} b)=n_{liq},
\ee

where $n_{gas}$ and $n_{liq}$ are the particle density of gas and liquid Argon, respectively; $T$ is the temperature, $2b=4\pi d^3/3$ is the covolume, that is, the excluded volume by the particle (here we take $d$ as a mean value of the viscosity radius and the gas radius) and $a=27T_c/64P_c$ is a measure of the particles attraction ($T_c=150.87\ K$, $P_c=4.898\ MPa$). Eq. (\ref{vanderwaals}) is a cubic equation in $n_{liq}$ which gives reasonably good results despite its simplicity. For this reason, we think that it is not necessary to derive a new state equation from the Helmholtz free energy.

Putting all together we are able to calculate the $\eta/s$ ratio 
in both liquid and gas states. The final result is plotted in Fig. 
\ref{arvoe} where a good agreement with the experimental data of 
\cite{CRC} is shown. 
One can see how  the KSS bound is maintained. 
Moreover, one can observe that for the liquid-gas phase transition 
$\eta/s$ presents a minimum and discontinuity at the phase transition 
(below the critical pressure, $P_c$). Above this pressure,
a minimum is still seen but the function is continuous. 

\newpage 
%%%%%%%%%%%%%%%%%%%%%%%%%%%%%%%%%%%%%%%%%%%%%%%%%%%%%%%%%%%%%%%%%
\section{Conclusions and outlook}
%%%%%%%%%%%%%%%%%%%%%%%%%%%%%%%%%%%%%%%%%%%%%%%%%%%%%%%%%%%%%%%%%
In this article we have argued, in agreement with previous authors, how it is likely that $\eta/s$ is a reasonable derived observable in relativistic heavy
ion collisions to pin down the phase transition and possible critical end
point between a hadron gas and the quark and gluon plasma/liquid.
We have contributed an evaluation
of the hadron-side $\eta/s$ that simultaneously encodes basic theoretical principles such as chiral symmetry and unitarity, and simultaneously produces a practical and good fit of the pion scattering phase shifts, by means of the Inverse Amplitude Method.
In so doing we have updated our past meson gas work. Our conclusions are in qualitative agreement with those of \cite{Chen:2007jq}. 

Since our lack of understanding of the non-perturbative dynamics
on the high-$T$ side of the phase transition to the quark-gluon phase
prevents us from matching asymptotic behavior of $\eta/s$ at high 
$T$ with the hadron gas, we have studied this KSS number in a related Sigma Model.
 We find numerically, and confirm with an 
analytical estimate, that keeping the $s$-channel amplitude one can 
isolate a minimum, and within reasonable calculational uncertainties, this coincides with the known phase transition of the model. A complete analysis is to be reported elsewhere.

Since we are not in possession of a good program that can proceed to
finite baryon density, we leave this for further investigation. Meanwhile
we have investigated the past observation that in going from a cross-over
to a first order phase transition, $\eta/s$ changes behavior, from 
having a continuous minimum to presenting a discontinuity. We choose, 
as very apt for theoretical study, atomic Argon. We employ standard
gas kinetic theory above the critical temperature and the Eyring
theory of liquids in the liquid phase. Whereas the discontinuity in $\eta/s$ is very clear for low pressures, theory and data are close
to matching (showing continuity) at high pressures where a crossover between
the two phases is seen in the phase diagram.

The conclusion is that indeed the minimum of the $\eta/s$  and the temperature of the phase transition might well be proportional. Whether the proportionality constant is exactly one could only be established by an exact calculation of the viscosity which is not theoretically at hand. 

As a consequence, we provide further theory hints to the currently proposed
method to search for the critical end point in hot hadron matter. If, as lattice
gauge theory suggests, a smooth crossover occurs between the hadron phase and
the quark-gluon phase, at least under the conditions in the Relativistic Heavy Ion
Collider where the baryon number is small at small rapidity, then one expects to see a minimum of viscosity over 
entropy density. In the FAIR experimental program however it might be possible
to reach the critical end point given the higher baryon density (since the energy per 
nucleon will be smaller), and whether the phase
transition is then first or second order can be inferred from the possibility of 
a discontinuity of $\eta/s$.

%%%%%%%%%%%%%%%%%%%%%%%%%%%%%%%%%%%%%%%%%%%%%%%%%%%%%%%%%%%%%%%%%%%%%%
\acknowledgments
We thank useful conversations and exchanges on $\eta/s$  with Jochen Wambach, Juan Maldacena, Dam Son, and Tom Cohen. 
This work has been supported by
grants FPA 2004-02602, 2005-02327,  BSCH-PR34/07-15875 (Spain)
%%%%%%%%%%%%%%%%%%%%%%%%%%%%%%%%%%%%%%%%%%%%%%%%%%%%%%%%%%%%%%%%%%%%5%

\end{document}